\documentclass[10pt, conference, compsocconf]{IEEEtran}

\usepackage{amsmath}
\usepackage{amssymb}
\usepackage{longtable}
\usepackage{cite}
\usepackage{graphicx}
\usepackage{epstopdf}
\usepackage{dcolumn}
\usepackage{stfloats}

\usepackage{color} 
\begin{document}

\title{Sublinear scaling of country attractiveness observed from Flickr dataset}

\author{
\IEEEauthorblockN{Iva Bojic}
\IEEEauthorblockA{SENSEable City Lab, MIT\\
Cambridge, MA, USA\\
Email: ivabojic@mit.edu}
\\
\IEEEauthorblockN{Stanislav Sobolevsky}
\IEEEauthorblockA{CUSP, New York University \\
Brooklyn, NY, USA\\
Email: sobolevsky@nyu.edu}
\and
\IEEEauthorblockN{Ivana Nizetic-Kosovic}
\IEEEauthorblockA{FER, University of Zagreb \\
Zagreb, Croatia\\
Email: ivana.nizetic@fer.hr}
\\
\IEEEauthorblockN{Vedran Podobnik}
\IEEEauthorblockA{FER, University of Zagreb \\
Zagreb, Croatia \\
Email: vedran.podobnik@fer.hr}
\and
\IEEEauthorblockN{Alexander Belyi}
\IEEEauthorblockA{SENSEable City Lab, SMART Centre\\
Singapore\\
Email: alex.bely@smart.mit.edu}
\\
\IEEEauthorblockN{Carlo Ratti}
\IEEEauthorblockA{SENSEable City Lab, MIT\\
Cambridge, MA, USA\\
Email: ratti@mit.edu}
}

\maketitle

\begin{abstract}
The number of people who decide to share their photographs publicly increases every day, consequently making available new almost real-time insights of human behavior while traveling. Rather than having this statistic once a month or yearly, urban planners and touristic workers now can make decisions almost simultaneously with the emergence of new events. Moreover, these datasets can be used not only to compare how popular different touristic places are, but also predict how popular they should be taking into an account their characteristics. In this paper we investigate how country attractiveness scales with its population and size using number of foreign users taking photographs, which is observed from Flickr dataset, as a proxy for attractiveness. The results showed two things: to a certain extent country attractiveness scales with population, but does not with its size; and unlike in case of Spanish cities, country attractiveness scales sublinearly with population, and not superlinearly.

\end{abstract}

\begin{IEEEkeywords}
human mobility; urban attraction; big data.

\end{IEEEkeywords}

\IEEEpeerreviewmaketitle

\section{Introduction}
\label{sec:introduction}

With the emergence of online social platforms (e.g., Twitter, Flickr, Facebook) not only do rich and famous people get attention, but also ordinary people can choose to participate in making their lives public. Moreover, even when people try to keep to themselves, their less concerned family members or friends can reveal a lot of things about their personal lives. Although we are still not fully aware to what extent will \textit{living online} affect/affects us, the huge data produced by humans in digital world represents almost an inexhaustible resource for different scientific studies (e.g., land use classification methods \cite{grauwin2015towards} or determining economical potential of cities \cite{sobolevsky2015predicting}).

The focus of this study is to investigate what catches people's eye using Flickr dataset of publicly available photographs and videos \cite{thomee2015new}. In our previous work we showed that city attractiveness in Spain scales superlinearly (exponent 1.5) with the city size \cite{Sobolevsky2015Scaling} where this high exponent denotes that for example attractiveness of a city three times bigger than the other one, is expected not to be three times, but on average five times higher. Moreover, this finding seems to be robust across different city definitions and also is true for two other datasets: Twitter and dataset of bank card transaction records. In this paper we extend our study to investigating how attractiveness scales on a country level. 

Various scaling studies have already been conducted on a city level as cities are important places that can transform a human life. While studies on socioeconomic parameters mostly show superlinear scaling laws, meaning that those parameters grow much faster than the city size, other studies on urban infrastructure reveal a sublinear relation to the city size. In that sense it has been shown that a bigger city boosts up human activity such as intensity of interactions \cite{schlapfer2012scaling}, creativity \cite{bettencourt2010urbscaling}, economic efficiency \cite{bettencourt2013origins}, as well as certain negative aspects: crime rates \cite{bettencourt2010urbscaling} or infectious diseases \cite{bettencourt2007growth}. On the contrary to the socioeconomic parameters, urban infrastructure shows sublinear scaling proving that cities are mostly built in an efficient way \cite{bettencourt2007growth, bettencourt2013origins}.

The novelty of our approach is not only in methods that we propose, but also in dataset that we use. In this paper we use publicly available Yahoo Flickr Creative Commons dataset, which is explained in more details in the next section, as a proxy for country attractiveness. Although this dataset was made publicly available only in 2014, scholars have already use it for building a user recommendation system for personalized tour based on their interests and points of interest visit duration \cite{lim2015personalized} or summarizing real-time multimedia events using Flickr as a source for social media \cite{shah2015eventbuilder}. However, to the best of our knowledge, nobody has use it to investigate country attractiveness seen through the lens of people taking photographs and videos.

The rest of the paper is organized as follows. Section~\ref{data_sets} introduces Yahoo Flickr Creative Commons dataset of more than 100 million media objects (i.e., photographs and videos). Section~\ref{attract} first shows results of aggregated country attractiveness scaling with country population and its size for countries in the whole world and then zooms in to countries grouped in different regions. Finally, Section~\ref{conclusion} concludes the paper and gives guidelines for future work.

\section{Dataset}
\label{data_sets}

In this study we use Yahoo Flickr Creative Commons dataset that is publicly available and can be downloaded from the following link: \cite{flickr_url}. This dataset is the largest public multimedia collection that has ever been released and was created as a part of the Yahoo Webscope program. It contains 100 million media objects, which have been uploaded to Flickr between 2004 and 2014 and published under a CC commercial or non-commercial license, of which approximately 99.2 million are photographs. Each object in the dataset is represented by its metadata: object identifier, user identifier, time stamp when it was taken, location (i.e., latitude and longitudinal coordinates) where it was taken (if available), and CC license it was published under. Additionally, the metadata for some objects also contains object title, user tags, machine tags and description, as well as a direct link for downloading the content.

\begin{figure}[t!]
  \centering
      \includegraphics[width=1\columnwidth]{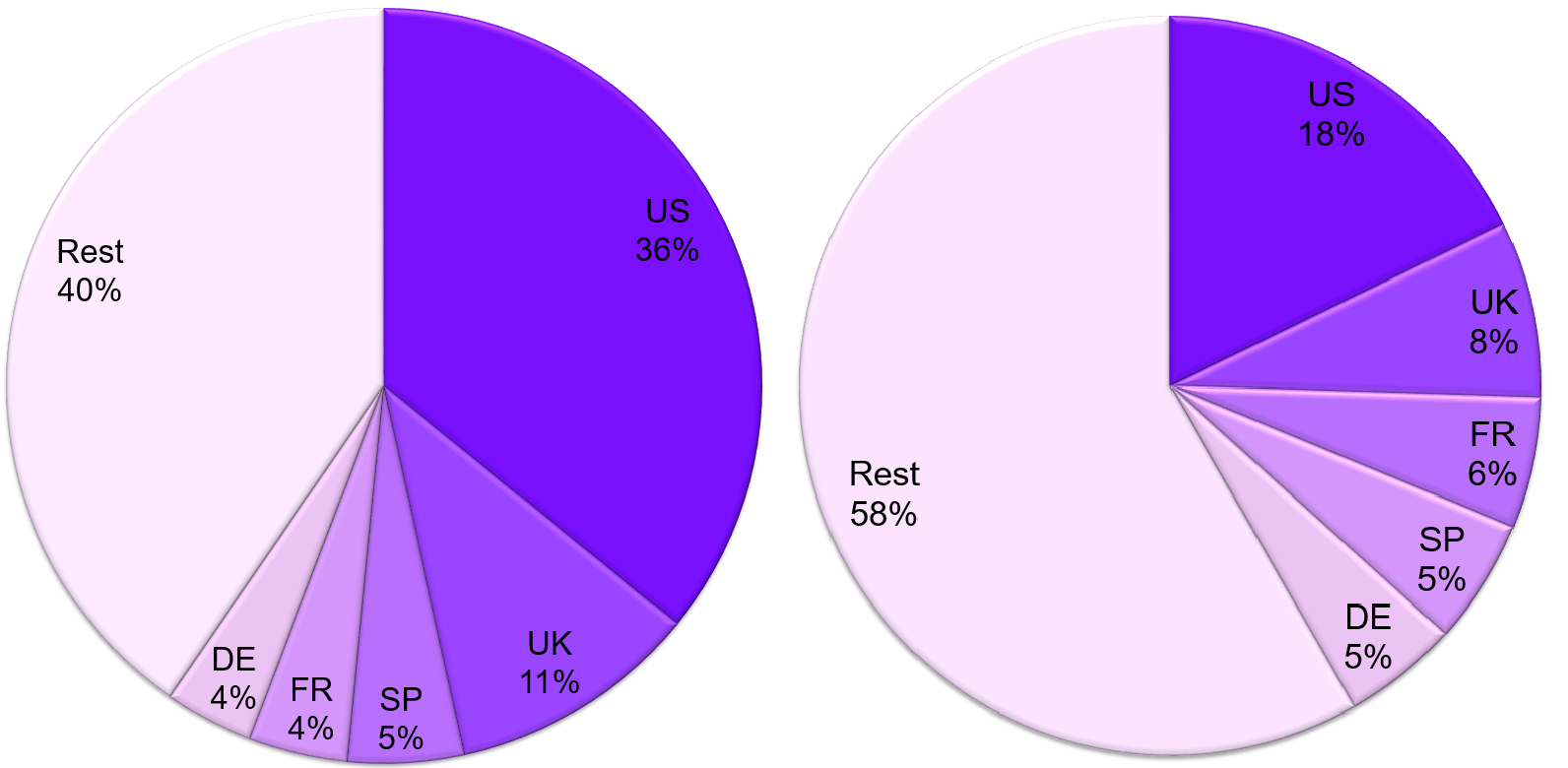}
    \caption{a) Distribution of media objects per country, b) distribution of unique users per country. Objects and users are not equally distributed since 60\% of media objects and more than 40\% of unique visitors belong to only five countries (i.e., US, UK, SP, DE, FR).}
 \label{fig:pie}
\end{figure}

Since the focus of the paper is to investigate how country attractiveness scales with the country size, we had to perform reverse geocoding so that every location (i.e., latitude and longitude coordinates) where media object was created is translated into a human readable format (i.e., country). But before applying a reverse geocoding method to the given dataset, we had to prune the data. In data pruning process we first omitted objects that were not geo-tagged (i.e., more than 50\%) and then objects that came with the wrong date format (652 in total). After both pruning and reverse geocoding we were left with 43,821,559 objects created in 242 countries around the world. Figure \ref{fig:pie} shows unequal distribution of unique users and media objects per country. 

However, the absolute numbers of media objects and unique visitors cannot be compared across different countries as countries are different in size and number of residents. When normalizing number of unique users with the number of people living in the country\footnote{Information about world population (i.e., number of people living in each country) was downloaded from the world bank website (http://data.worldbank.org/indicator/SP.POP.TOTL). Since Flickr dataset includes media objects created between years 2004 and 2014, in our normalization process we do not use the population for a specific year, but calculate the average number of people living in the country for the same period of ten years.}, we get completely different order of first five countries starting with Iceland where one Flicker user comes on 180 of its residents, followed by Greenland, San Marino, Andorra and Cayman Islands. With the exception of Iceland, this normalization shows a bias towards small countries that are touristically attractive. Moreover, when normalizing the number of media objects with the country size\footnote{Country's total area was also downloaded from the world bank website (http://data.worldbank.org/indicator/AG.LND.TOTL.K2) and was prepared in the same way as population, i.e., calculating an average value for the time period between 2004 and 2014.} we get similar order of the first five countries: Iceland, Greenland, Cayman Islands, United States Virgin Islands and Palau. 

\begin{figure}[h!]
  \centering
      \includegraphics[width=1\columnwidth]{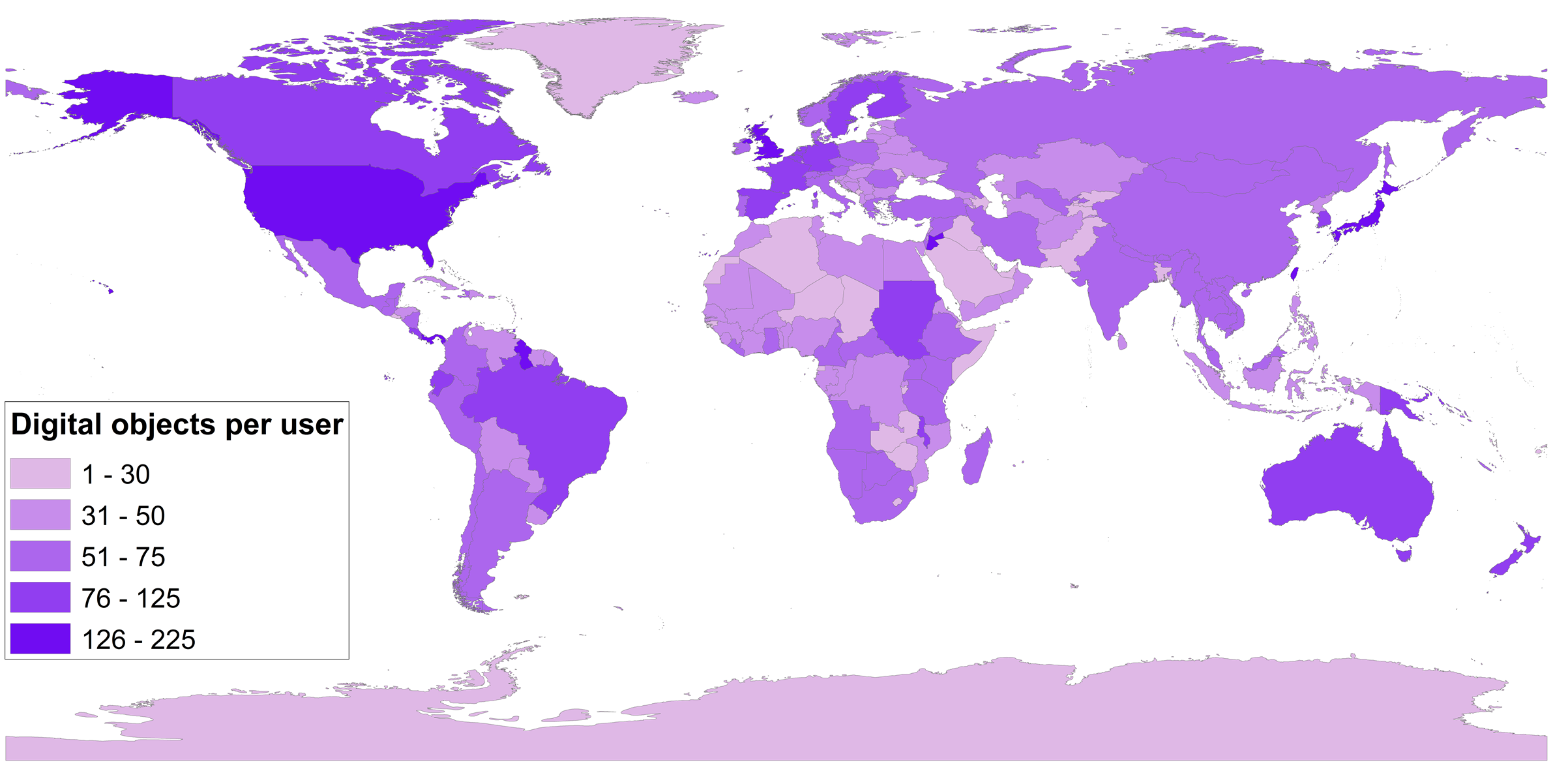}
  \caption{Map showing an average number of media objects produced by users in a specific country. The first five countries are as follows: Taiwan, United States, Panama, Guyana, Japan with 222, 215, 197, 174 and 162 media objects per user respectively.}
 \label{fig:map}
\end{figure}

Finally, Figure \ref{fig:map} shows an average number of media objects that every user produces in a given country ranging from more than 200 objects per user made in Taiwan and the United States to less than 10 made in for example Chad and San Marino.

\section{Scaling of country attractiveness}
\label{attract}

Rather than including all human activity captured in the dataset, for the purpose of investigating country attractiveness scaling laws, we only included foreign users and their activity. Since Flickr dataset does not include information about user home countries, we used the same methodology as we used in our previous papers \cite{Paldino2015Urban, bojic2015choosing} to infer where people live. We say that a particular user lives in a particular country if there he/she made the maximum number of media objects and spend the maximal number of days. Then the activity of those users for whom we detected their home countries is used when calculating aggregated country attractiveness in two different ways: number of media objects taken by foreign users compared to the country size and its population. Aggregated country attractiveness denotes that we included all media objects without taking into consideration time when they were made.

\begin{figure*}[b!]
\centering
\includegraphics[width=.45\textwidth]{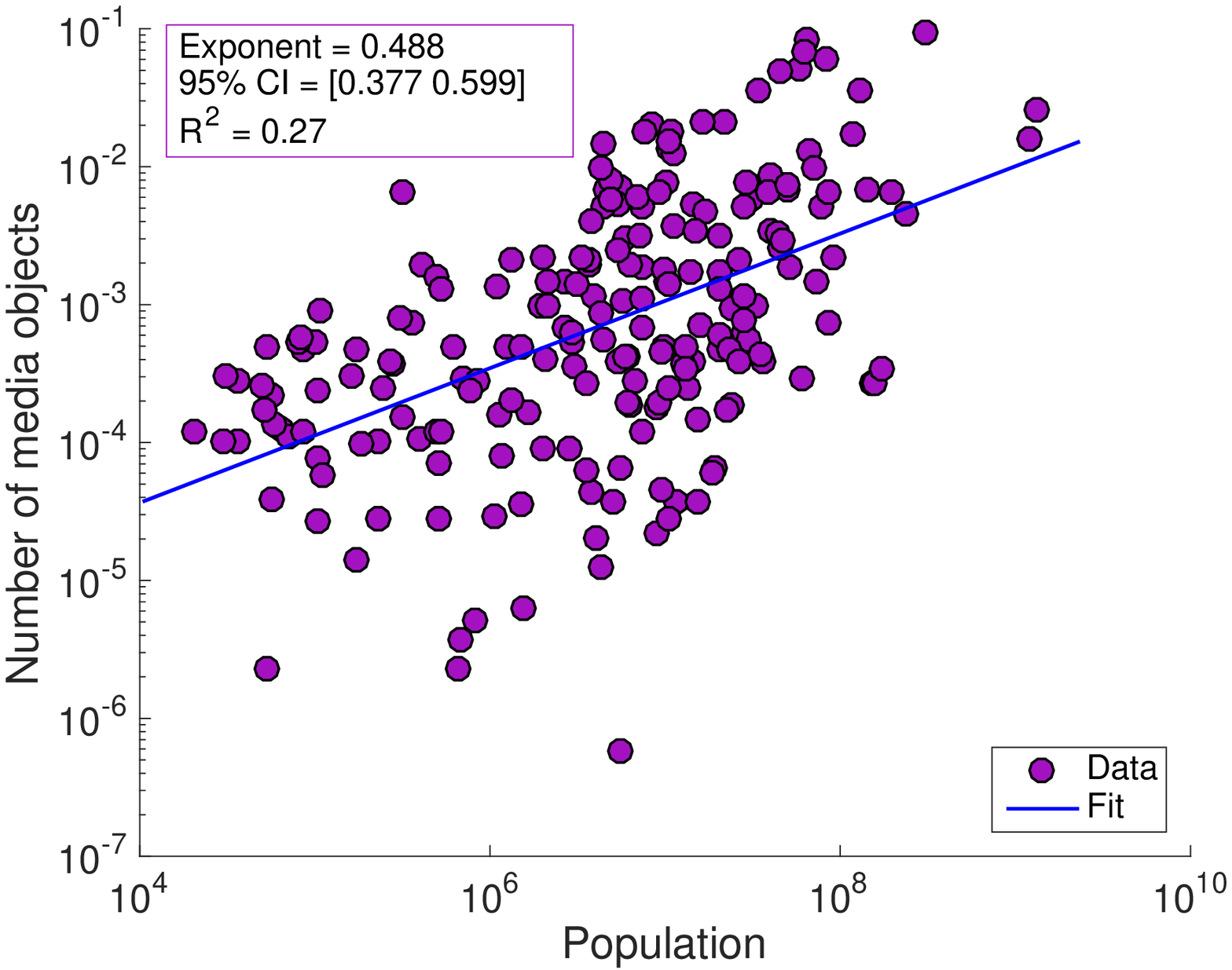}
\includegraphics[width=.45\textwidth]{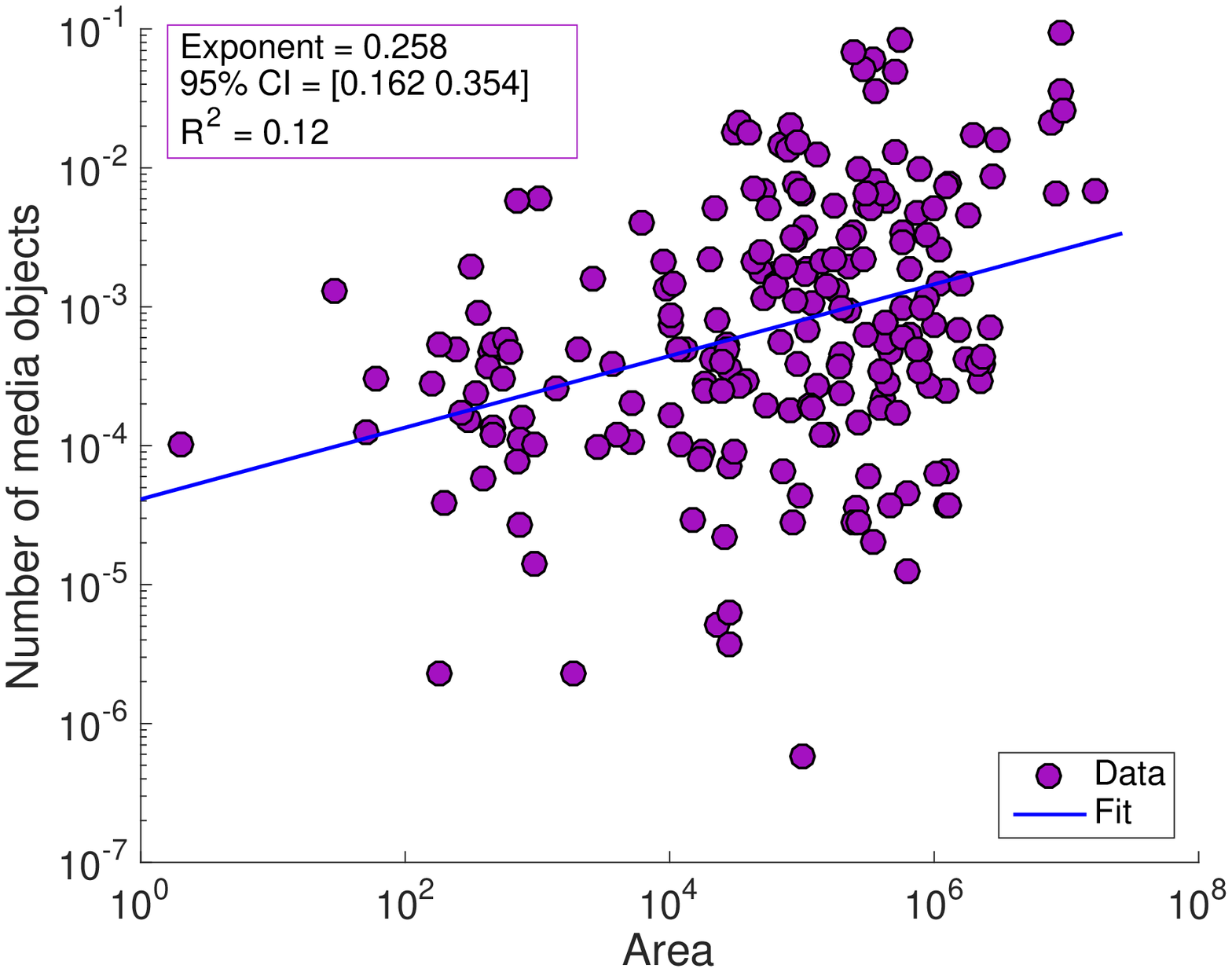}
\caption{\label{fig::country_scaling}Scaling of aggregated county attractiveness where X-axis represents the number of people living in the country or the country size, while Y-axis represents the fraction of the number of media objects made in a particular country versus the total amount of objects made in all countries.}
\end{figure*}

Figure \ref{fig::country_scaling} shows the results when fitting a power-law dependence $A\sim a p^{\beta}$ or $A\sim a s^{\beta}$ to the calculated number of media objects denoting aggregated country attractiveness $A$ and population $p$ or size $s$. Fitting is performed on a log-log scale where it becomes a simple linear regression problem $log(A)\sim log(a)+\beta\cdot log(p)$ or $log(A)\sim log(a)+\beta\cdot log(s)$. As proposed in \cite{bettencourt2007growth} scaling can fall in three different categories depending on the value of parameter $\beta$: linear ($\beta = 1$), sublinear ($\beta < 1$) and superlinear ($\beta > 1$). Unlike city attractiveness \cite{Sobolevsky2015Scaling}, country attractiveness for countries all around the world scales sublinearly with both country population $p$ and size $s$. A measure of goodness-of-fit of linear regression $R^{2}$ is $26.7\%$ for population $p$ and $11.9\%$ for size $s$, denoting that sublinear scaling trend of country attractiveness with the population is more emphasized than scaling trend with country size. However, values for both $R^{2}$ are small enough meaning that we cannot get a perfect fit for all counties in the world. Therefore, we divided world countries into eleven regions\footnote{Data used in this study can be downloaded from the following link: http://www.statvision.com/webinars/Countries\%20of\%20the\%20world.xls.} and checked if scaling laws are more present in certain regions rather than all world. 

\begin{table}[t!]
\centering
\caption{World regions with number of countries belonging to them, population, scaling coefficient $\beta$ and parameter $R^{2}$.}
\label{tab::region_scaling}
\begin{tabular}{|l|r|r|r|}
\hline
\begin{tabular}[c]{@{}l@{}}Region name \\ (no of countries)\end{tabular} & Population \  & $\beta$\ \ \ \ & $R^{2}$ (\%)    \\ \hline
Northern America (4)                                                     & 340,100,983   & 0.777	 	& 98.5 \\ \hline
Western Europe (25)                                                      & 409,402,265   & 0.715		& 93.8 \\ \hline
Baltics (3)                                                              & 6,607,817     & -0.499    	& 89.0 \\ \hline
Northern Africa (6)                                                      & 161,701,943   & 0.964	   	& 75.4 \\ \hline
\begin{tabular}[c]{@{}l@{}}Commonwealth \\
of Independent States (12)\end{tabular}									 & 280,892,606   & 0.933		& 63.0 \\ \hline
Oceania (16)                                                             & 35,557,584    & 0.850	   	& 59.2 \\ \hline
\begin{tabular}[c]{@{}l@{}}Latin America and \\
Caribbean (40)\end{tabular} 											 & 589,003,980   & 0.466        & 57.6 \\ \hline
Eastern Europe (12)                                                      & 114,521,789   & 0.778		& 40.7 \\ \hline
Sub-Saharan Africa (48)                                                  & 836,231,137   & 0.638	   	& 27.6 \\ \hline
Asia (ex. Near East) (25)                                                & 3,787,575,787 & 0.363	   	& 25.7 \\ \hline
Near East (14)                                                           & 205,611,734   & 0.344	   	& 11.0 \\ \hline
\end{tabular}
\end{table}

\begin{figure*}[t!]
\centering
\includegraphics[width=.45\textwidth]{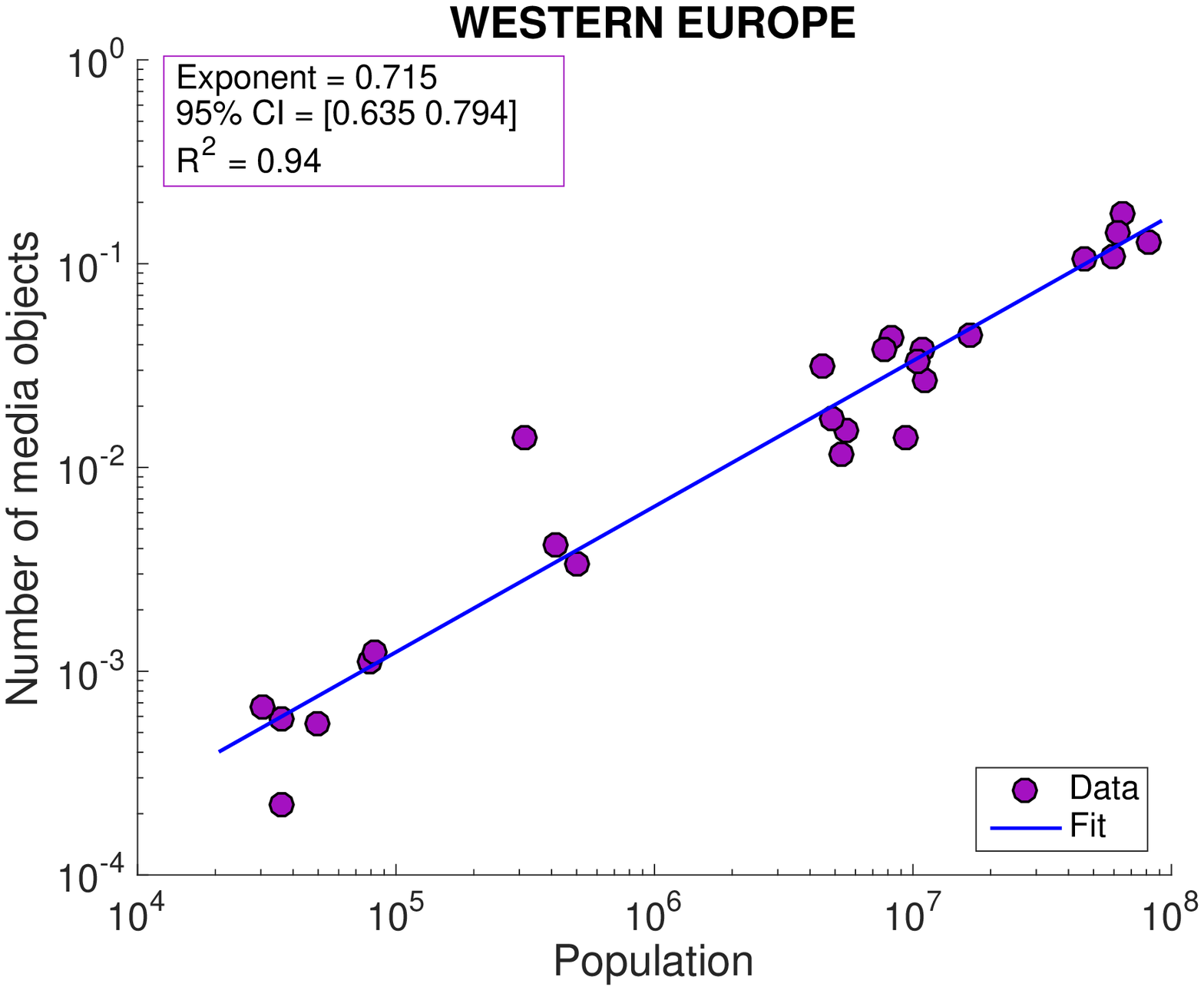}
\includegraphics[width=.45\textwidth]{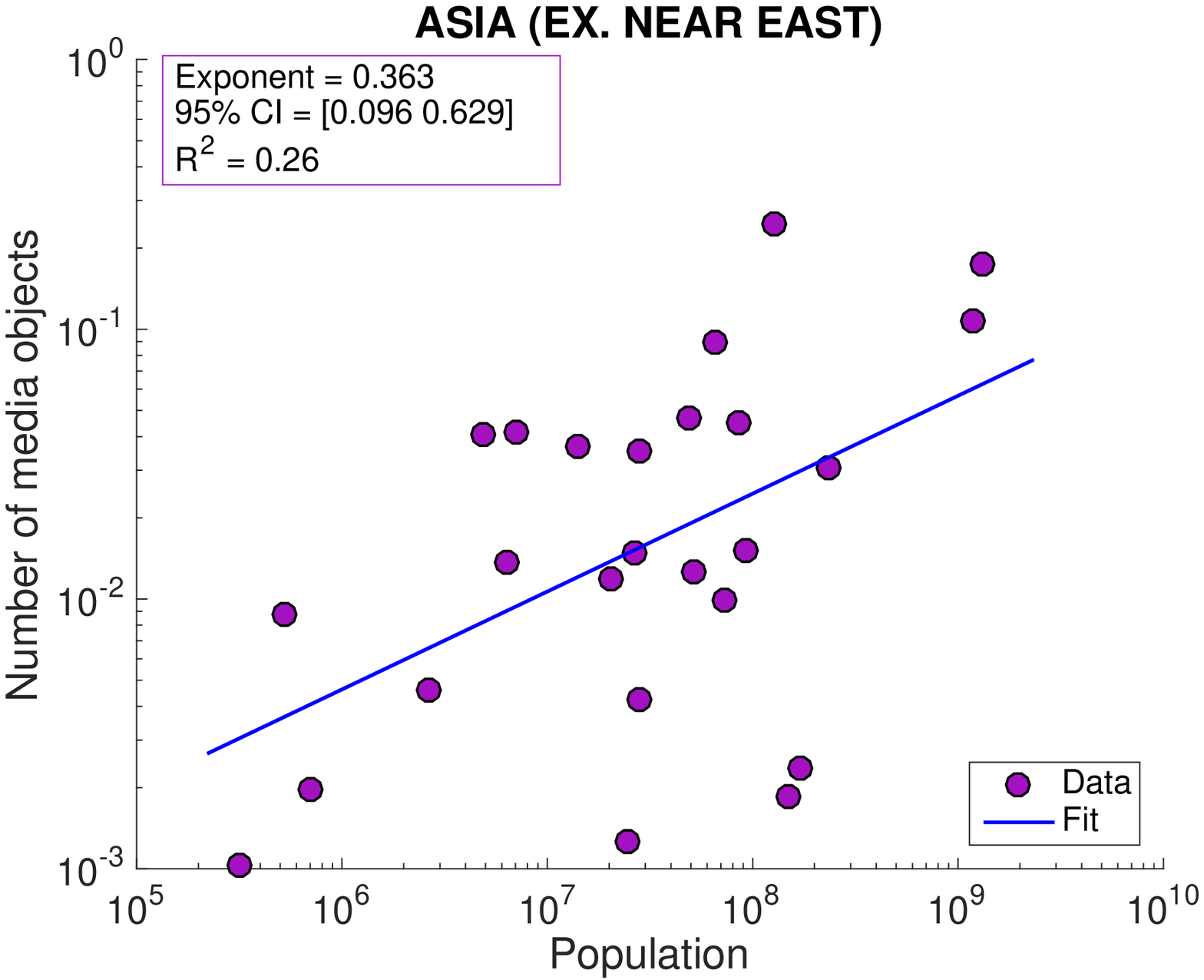}
\caption{\label{fig::region_scaling}Scaling of aggregated county attractiveness for Western Europe (on left) and Asia excluding Near East (on right).}
\end{figure*}

Table \ref{tab::region_scaling} shows that all countries belonging to different regions are sublinearly scaling with their population. Regions are sorted descending by values of $R^{2}$, i.e., best fit. Results show that almost a perfect fit (almost 99\%) is for region called Northern America with only four countries, followed by Western Europe which consists of 25 countries. However, the number of countries in the region is not proportional with a measure of goodness-of-fit of linear regression $R^{2}$ as regions Asia (ex. Near East) and Near East have the same number of countries as Western Europe or less and their fit is less than 26\%. Aggregated country attractiveness for Western Europe and Asia (ex. Near East), which both are composed of 25 countries, is shown in Figure \ref{fig::region_scaling}.

Since there were not obvious correlations between $R^{2}$ and parameters form Table \ref{tab::region_scaling}, we calculated the  correlation coefficient for $R^{2}$ and country area, population, GDP, density, coastline and urban population. The correlation coefficient, which is a value between $-1$ and $+1$, tells how strongly two variables are related to each other. The smallest correlation coefficient was between $R^{2}$ and density (-0.102), while the biggest was with GDP (0.678).

\section{Conclusions}
\label{conclusion}
In this study we analyzed aggregated country attractiveness in dependence with its population and size for countries around the world. The results on country attractiveness showed that there is not a direct correlation between it and country size, unlike with population where different regions of the world showed different behavior. Namely, regions like Western Europe, Northern America and Baltics showed linearity on log-log scale, while regions like Sub-Saharan Africa, Asia and Near East are not well fitted. 

Further, we also investigated correlations between the goodness-of-fit $R^{2}$ and various parameters of the countries (e.g., density, coastline) and discovered the most significant correlation with country GDP, i.e., 0.678. Regardless to the goodness-of-fit, all regions showed sublinear scaling with their population, rather than superlinear one discovered within prior studies on aggregated city attractiveness. 

\section*{Acknowledgments}
The authors would like to thank BBVA, MIT SMART Program, Center for Complex Engineering Systems (CCES) at KACST and MIT, Accenture, Air Liquide, The Coca Cola Company, Emirates Integrated Telecommunications Company, The ENELfoundation, Ericsson, Expo 2015, Ferrovial, Liberty Mutual, The Regional Municipality of Wood Buffalo, Volkswagen Electronics Research Lab, UBER, and all the members of the MIT Senseable City Lab Consortium for supporting the research. Part of this research was also funded by the research project "Managing Trust and Coordinating Interactions in Smart Networks of People, Machines and Organizations", funded by the Croatian Science Foundation.

\bibliography{literature}

\end{document}